\documentclass[12pt]{article}

\usepackage{amsmath}
\usepackage{amssymb}
\usepackage{multirow,booktabs}
\usepackage{graphicx}
\usepackage{cite}
\usepackage{hyperref}
\usepackage{color} 
\usepackage{color, colortbl}
\definecolor{LightCyan}{rgb}{0.88,1,1}
\definecolor{Gray}{gray}{0.94}
\usepackage{setspace} 
\usepackage[labelfont=bf,labelsep=period,justification=raggedright]{caption}
\makeatletter
\renewcommand{\@biblabel}[1]{\quad#1.}
\makeatother

\date{\today}

\begin{document}

\begin{flushleft}
{\Large
\textbf{Distance to the scaling law: a useful approach for unveiling relationships between crime and urban metrics}
}
\\
Luiz G. A. Alves,
Haroldo V. Ribeiro$^{\ast}$,
Ervin K. Lenzi,
Renio S. Mendes
\\
Departamento de F\'isica and National Institute of Science and Technology for Complex Systems, Universidade Estadual de Maring\'a, Maring\'a, PR 87020-900, Brazil
\\
$^{\ast}$ E-mail: hvr@dfi.uem.br
\end{flushleft}

\section*{Abstract}
We report on a quantitative analysis of relationships between the number of homicides, population size and other ten urban metrics.
By using data from Brazilian cities, we show that well defined average scaling laws with the population size emerge when investigating 
the relations between population and number of homicides as well as population and urban metrics. We also show that the fluctuations 
around the scaling laws are log-normally distributed, which enabled us to model these scaling laws by a stochastic-like 
equation driven by a multiplicative and
log-normally distributed noise. Because of the scaling laws, we argue that it is better to employ logarithms in order to describe
the number of homicides in function of the urban metrics via regression analysis. In addition to the regression analysis, we propose
an approach to correlate crime and urban metrics via the evaluation of the distance between the actual value of the number of
homicides (as well as the value of the urban metrics) and the value that is expected by the scaling law with the population size. This approach have proved
to be robust and useful for unveiling relationships/behaviors that were not properly carried out by the regression analysis, such as 
$i)$ the non-explanatory potential of the elderly population when the number of homicides is much above or much below the scaling law, $ii)$ the fact that
unemployment has explanatory potential only when the number of homicides is considerably larger than the expected by the power law,
and $iii)$ a gender difference in number of homicides, where cities with female population below the scaling law are characterized by a number 
of homicides above the power law.

\section*{Introduction}
The study of social complex systems has been the focus of intense research in the last decades~\cite{amaral,castellano,conte}. Elections~\cite{epl,Chatterjee}, population growth~\cite{rybski1,rybski2}, economy~\cite{amaral2,mantegna,peron}, and language~\cite{wichmann,petersen,luciano} are just a few examples of social activities that have been recently investigated.  Such investigations are expected to provide a better understanding of how our society is organized and also to point out better strategies for resource management, service allocation, and political strategies. In this social context, crime is one of the most worrying activity for our society and to understand and to prevent crime acts is a huge challenge~\cite{kates,iglesias1,iglesias2}. Moreover, since nowadays more than a half of the human population lives in cities~\cite{crane,wri}, it is crucial to analyze possible connections between criminality and urban metrics.  

In fact, there exist several works that point out relationships between the number of crime acts and urban indicators such as income, unemployment and inequality~\cite{becker,ehrlich1,ehrlich2,glaeser1,glaeser2}. Most of these papers employ regression analysis, where the dependent variable is the crime indicator (usually  the number of a particular crime act) and the independent variables are urban indicators~\cite{blau,bailey,kennedy,kelly,levitt1,levitt2,hojman1,hojman2,sachsida,poveda}. However, most of these studies does not take into account the functional form of the relationships between crime, urban indicators and the population; usually assuming these relationships to be linear~\cite{gordon}. On the other hand,  several works have shown that crime and urban indicators obey scaling laws with the population size of the cities and also between themselves~\cite{pnas,epjb,nature,plos,lievano}. For instance,  the number of homicides grows super-linearly with the population~\cite{lievano,alves}. Do not consider these scaling laws may be one of the reasons that several regression-based analysis led to controversial conclusions~\cite{gordon}. Furthermore, if we assume that these scaling laws with the population size are somehow a natural expression of how cities are organized, accounting for the scaling phenomenon is also very
important for achieving a fairer comparison between cities with different population sizes. 

Here we investigate a procedure that may help to solve this problem. The approach consists of defining a ``distance'' between the crime or urban indicators and the main tendency expected by the scaling laws with the population size. This approach is based on the recent idea of relative competitiveness proposed by Podobnik \textit{et al.}~\cite{podobnik}  in the economic context. Our paper is thus organized as follows. We start by presenting our data of urban and crime indicators of Brazilian cities and also an intensive characterization of the scaling laws existing between these indicators and the population size. We also employ a linear regression model for explaining the number of crime acts (homicides) in terms of the urban indicators. Next, we use the previously-discussed distance in an attempt to investigate relationships/patterns between crime and urban metrics that do not appear in the regression analysis. Finally, we present a summary of our results.

\section*{Materials and Methods}

\subsection*{Data presentation}
We have accessed data of the Brazilian cities in the year of 2000 made freely available by the Brazil's public healthcare system --- DATASUS~\cite{datasus}.
These data are also attached to our paper in the \textit{Supplementary Table S1}. Here, despite there being other definitions~\cite{angel}, we have considered that cities are the smallest administrative units with a local government 
and it is not our intention to discuss the role of other definitions. The data consist of the population size ($N$) and the number of homicides ($H$) as well as
ten urban indicators ($Y$) at city level: number of cases of child labour, elderly population size (older than 60 years), female population size, 
gross domestic product (GDP), GDP per capita, number of illiterate (older than 15 years), average family income, male population size, number of sanitation facilities, and number of unemployed (older than 16 year). More details about urban indicators can be found in the \textit{Supplementary Text S1}. 
Observe that we have chosen the number of homicides as our crime indicator. This is a widely used choice~\cite{lievano} due the fact that homicide data
are more reliable, since this ultimate expression of violence is almost always reported. Also, our ten urban indicators are usually listed as crime 
determinants~\cite{gordon}. Furthermore, we have considered only cities with at least one case of homicide in our analysis.

\section*{Results and Discussion}

\subsection*{Scaling laws between crime, urban metrics and population}

\begin{figure}[!ht]
\includegraphics[scale=0.4]{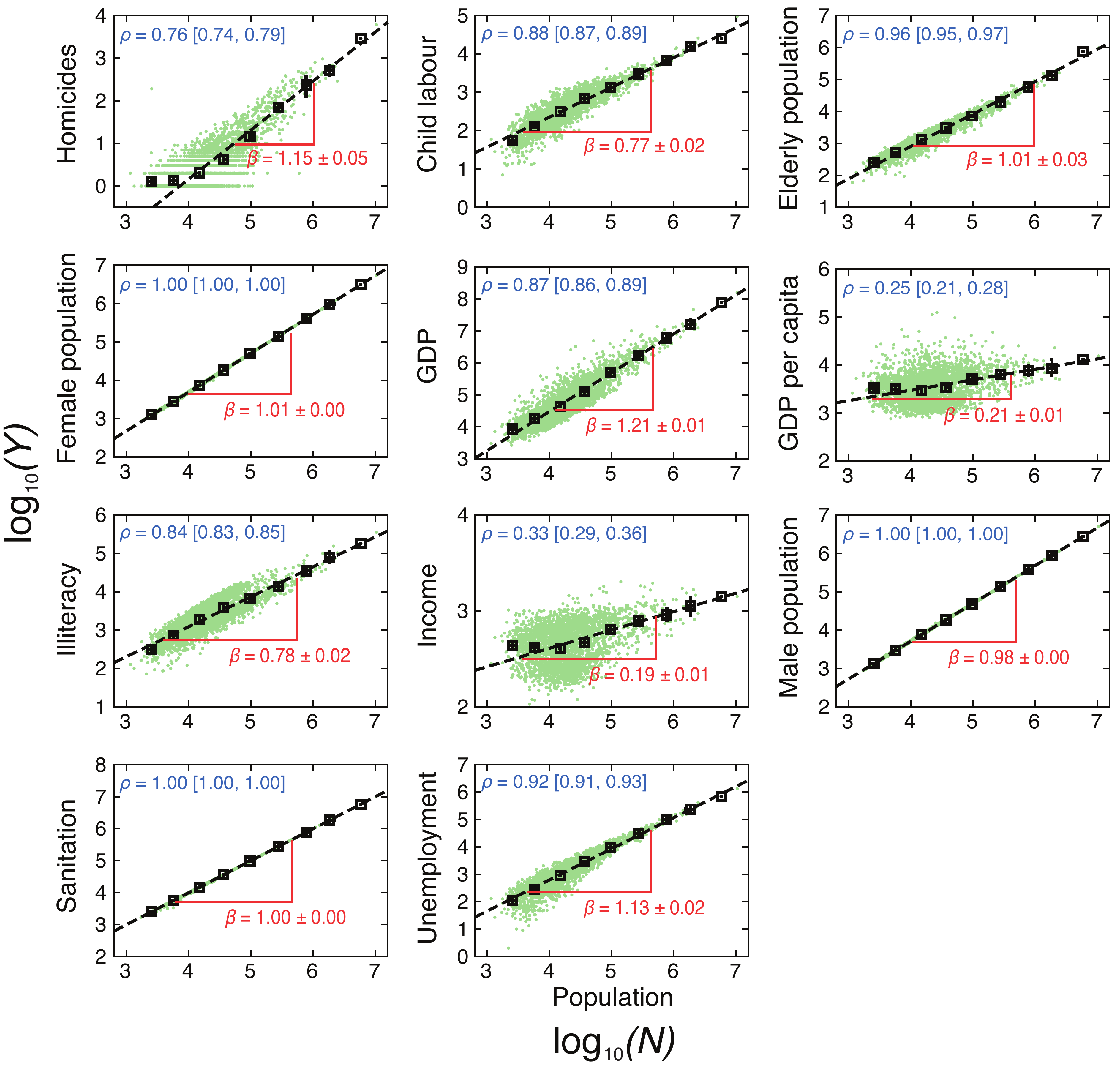}
\caption{
{\bf Scaling laws between the population size and the urban indicators.} 
In each plot, the green dots are base-10 logarithmic of the values of the urban indicator ($Y$) versus the population size ($N$) for a given city. 
The black squares are average values of the data binned in 10 equally spaced windows and the error bars are 95\% confidence intervals for these average values obtained via bootstrapping~\cite{efron}. The values of the Pearson correlation coefficients $\rho$ (as well as the 95\% confidence intervals) of these relationships are shown in each plot. The straight dashed lines are linear fits (by least square method) to the average relationships and the slope of these lines are equal to the power law exponent $\beta$ (shown in each plot). 
}
\label{fig:1}
\end{figure}

We start by revising the question of whether homicides and urban metrics present scaling relations with the population 
size (see also Refs.~\cite{pnas,epjb,nature,plos,lievano,alves}). 
For the sake of simplicity, let us denote the population size by $N$ and the urban indicators by $Y$. We thus want to check if $Y$ is a power law
function of $N$, that is, $Y\sim N^\beta$, where $\beta$ is the power law exponent. Figure~\ref{fig:1} shows a scatter plot of $\log_{10}Y$ versus $\log_{10}N$
for all urban indicators, starting with the number of homicides and passing through all the ten urban metrics. We note that, despite the existence 
of considerable noise in some relationships, the scaling laws with the population size are perceptible. In order to overcome the noise and uncover
the main tendency in these relationships, we have binned the data in $w$ windows equally spaced in $\log_{10}N$ and evaluated the average values of
the points within each window. The square symbols shown in Fig.~\ref{fig:1} represent these average values and the dashed lines are linear fits.
Note that linear functions describe quite well all the average relations, that is, the equation
\begin{equation}
\langle \log_{10} Y \rangle_{w} = A+\beta \log_{10} N \,
\end{equation}
holds for all the urban indicators. Here, $\langle \log_{10} Y \rangle_{w}$ is the average value of $Y$ within each one of the $w$ windows, $A$ is a constant
and $\beta$ is the power law exponent (shown in Fig.~\ref{fig:1}). We have thus confirmed that there are scaling laws between the average values of the
urban indicators $Y$ and the population $N$. It is worth to remark that these average relationships are very robust when
varying the number of windows $w$ (see Fig.~\ref{sfig:1}).

\begin{figure}[!ht]
\includegraphics[scale=0.4]{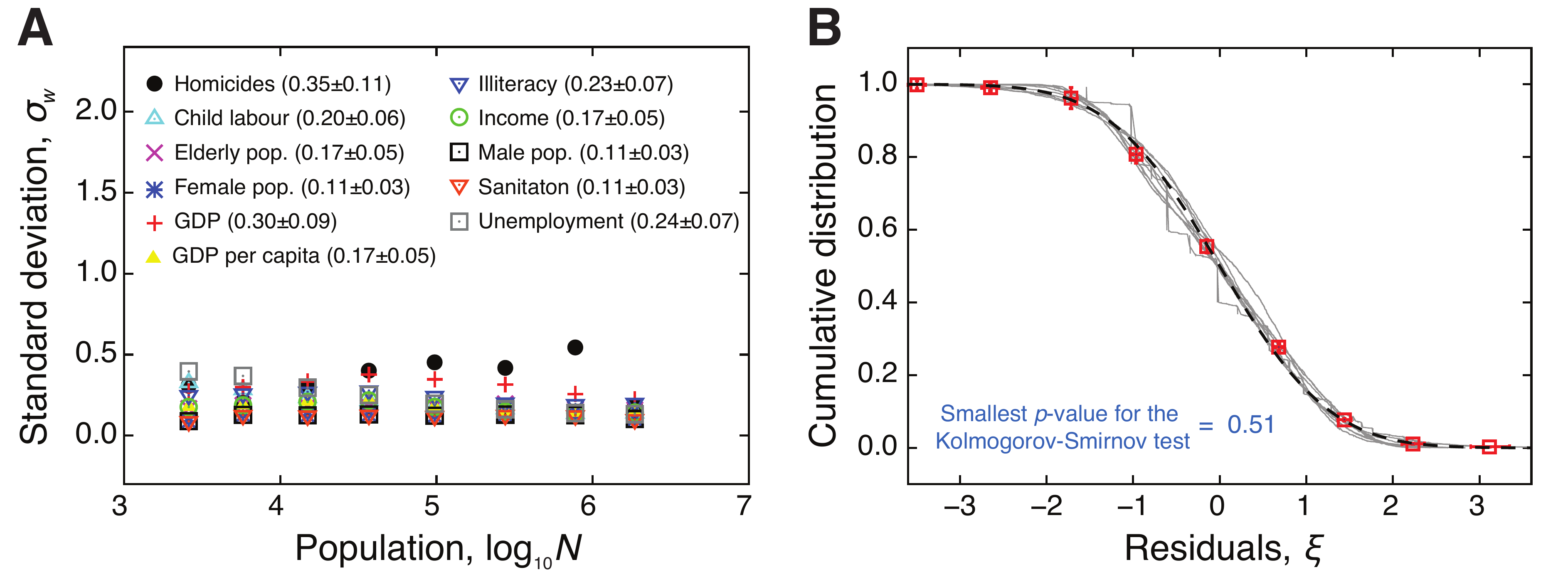}
\caption{
{\bf Fluctuations around the scaling laws.}
(A) Standard deviation $\sigma_w$ of the fluctuations around the scaling laws (in base-10 logarithmic scale) in each one of the 
$w=10$ equally spaced windows. We note that the standard deviation is almost a constant function of the population 
for all urban indicators. The average value of $\sigma_w$ over the population windows are
shown in the plot legends. (B) Cumulative distributions of the normalized fluctuations $\xi$ around the scaling laws. In this plot, each gray line is a
distribution for a given indicator, the squares are the average values of these cumulative distributions and the error bars are 95\% confidence 
intervals obtained via bootstrapping~\cite{efron}. We note that the Gaussian distribution (dashed line) describes quite well these distributions. 
In particular, the smallest $p$-value of the Kolmogorov-Smirnov tests is $0.51$, showing that we cannot reject the normality of the fluctuations.
}
\label{fig:2}
\end{figure}

Another striking feature of Fig.~\ref{fig:1} is the fluctuation around the power law tendency. We have observed that the standard deviation
\begin{equation}
\sigma_w = \sqrt{\langle (\langle \log_{10} Y \rangle_{w} - \log_{10} Y)^2 \rangle_{w}}
\end{equation}
within each window practically does not change with the population size $N$ for all urban indicators (Fig.~\ref{fig:2}A). We have also verified that
the normalized residuals around the power law,
\begin{equation}
\xi = \frac{\log_{10} Y(N) - \langle \log_{10} Y \rangle_{w} }{\sigma_w}\,,
\end{equation}
are normally distributed with zero mean and unitary standard deviation (Fig.~\ref{fig:2}B). In particular, the Kolmogorov-Smirnov test~\cite{corder} 
cannot reject the normality of $\xi$ for all the urban indicators (the $p$-values are all larger than 0.51).

Our previous analysis thus enable an elegant formulation to the average scaling laws and also to the noise around these tendencies. Mathematically, we can write
\begin{equation}\label{eq:powerlaw}
Y=\mathcal A\, \eta(N)\, N^\beta
\end{equation}
or, equivalently 
\begin{equation}
\log_{10} Y= \log_{10} \mathcal A+ \beta \log_{10}N + \log_{10} \eta (N)\,,
\end{equation}
where $\log_{10} \mathcal A = A$ and $\log_{10} \eta(N)=\sigma_w\, \xi(N)$. Notice that, since $\xi(N)$ is normally distributed, $\eta (N)$
should be distributed according to a log-normal distribution. In addition to describe the average scaling laws, Eq.~(\ref{eq:powerlaw})
represents a stochastic-like process where the urban indicator $Y$ follows a power law relation with the population $N$ driven by a multiplicative 
noise log-normally distributed.

\subsection*{Regression model: homicides versus urban metrics}

As we have mentioned in the introduction, a considerable part of the literature about criminality tries to correlate crime indicators to other urban metrics.
Usually, these relationships are obtained from linear regression models, despite the explicit nonlinearities present in these 
variables such as the previous scaling laws. In this context, it is not uncommon to observe linear regression-based analysis leading to controversial conclusions~\cite{gordon}.
A simple alternative that may overcome these nonlinearities is to employ the logarithmic of the variables, that is,
\begin{equation}\label{eq:regression}
\log_{10} H(i)= C_0 + \sum_{k} C_k \log_{10} Y_k(i) + \epsilon(i)\,.
\end{equation}
Here, $H(i)$ is the number of homicides in the city $i$, $Y_k (i)$ is the $k$-th ($k>1$) urban indicator of the city $i$, 
$C_0$ is the intercept coefficient, $C_k$ ($k>1$) is the linear coefficient that quantifies the explicative effect of $\log_{10} Y_k (i)$, 
and $\epsilon(i)$ is the noise term accounting for the effect of unmeasurable factors.

\begin{table}[!ht]
\caption{
{\bf Regression model coefficients.} Values of the linear coefficients $C_k$ obtained via ordinary least-squares fits with a correction to heteroskedasticity.
Here, $t$ is the value of the $t$-statistic and $p$ is the two-tail $p$-value for testing the hypothesis that the coefficient $C_k$ is different from zero.
}
\renewcommand{\arraystretch}{1.2}
\centering
\begin{tabular}{clcrcr}
\toprule
$k$ & Indicator $Y_k$ & Coefficient $C_k$ & Standard Error & $t$ & $p>|t|$  \\
& & 95\% Confidence Interval &  &  &   \\
\midrule
\rowcolor{Gray}
0 & Intercept & 322.932 & 84.653 & 3.81 & 0.000 \\
\rowcolor{Gray}
 &  & [156.944, 488.920] &  &  &   \\
1 & Child labour & -0.146 & 0.035 & -4.11 & 0.000 \\
 &  & [-0.216, -0.076] &  &  &   \\
\rowcolor{Gray}
2 & Elderly population & -0.647 & 0.066 & -9.81 & 0.000 \\
\rowcolor{Gray}
 &  & [-0.777, -0.518] &  &  &   \\
3 & Female population & -56.644 & 15.488 & -3.66 &  0.000\\
 &  & [-87.015, -26.274] &  &  &   \\
\rowcolor{Gray}
4 & GDP & 121.127 & 31.375 & 3.86 & 0.000 \\
\rowcolor{Gray}
 &  & [59.605, 182.648] &  &  &   \\
5 & GDP per capita & -120.987 & 31.375 & -3.86 & 0.000 \\
 &  & [-182.509, -59.465] &  &  &   \\
\rowcolor{Gray}
6 & Illiteracy & 0.213 & 0.051 &  4.11 &  0.000 \\
\rowcolor{Gray}
 &  & [0.111, 0.314] &  &  &   \\
7 & Income & 0.223 & 0.073 & 3.05 & 0.002\\
 &  & [0.079, 0.367] &  &  &   \\
\rowcolor{Gray}
8 & Male population & -62.459 & 16.068 & -3.89 &  0.000 \\
\rowcolor{Gray}
 &  & [-93.967, -30.952] &  &  &   \\
9 & Sanitation & -0.665 & 0.929 & -0.72 & 0.474  \\
 &  & [-2.487, 1.156] &  &  &   \\
\rowcolor{Gray}
10 & Unemployment & -0.026 & 0.028 & -0.94 & 0.347 \\
\rowcolor{Gray}
 &  & [-0.082, 0.028] &  &  &   \\
\midrule
\multicolumn{6}{r}{Adjusted $R^2=0.62$}\\
\bottomrule
\end{tabular}
\label{tab:model}
\end{table}

We have applied the previous model to our data by using ordinary least-squares fit with a correction to heteroskedasticity~\cite{davidson} and 
the results are summarized in Table~\ref{tab:model}. We first note that, except
for sanitation and unemployment, all the urban indicators have explanatory potential for describing the number of homicides. Also,
the value of the adjusted $R^2$ points out that the model account for about 62\% of the observed variance in number of homicides. 
When analyzing the individual effects of the urban indicators, we note that child labour, elderly population, female population, GDP per
capita, and male population are negatively correlated with the number of homicides ($H$ decreases with the increasing of these indicators).
On the other hand, GDP, illiteracy, and income are positively correlated with the number of homicides ($H$ increases with the increasing of 
these indicators). Despite the lack of a more adequate comparison with our data, our regression results agree but also disagree
with some empirical findings of the criminology literature. For instance, we have found that there is no statistically significant correlation
between unemployment and homicides, while a positive and statistically significant correlation between illiteracy and homicides was observed.
However, these indicators are among those leading to controversial conclusions, as pointed out by Gordon~\cite{gordon}. 

Naturally, our regression model is quite simple and several improvements are possible. For instance, some of these metrics may display correlations
and, consequently, one metric may affect the predicability of another, a phenomenon known as mediation~\cite{preacher}. A possible
manner for reducing this effect is by combining some of the metrics and running different regression models. Another possibility is
to employ principal component analysis (PCA) for reducing redundancy among the urban metrics. Nevertheless, other problems such as bias
in the selection of urban metrics and difficulties in drawing qualitative conclusions in terms of the PCA axis are still present. 
Here, instead of discussing the possible controversies that Table~\ref{tab:model} may exhibit as well as possible manner of improving our
regression results, we will compare this simple regression analysis with our new approach based on the deviations of the scaling laws.

\subsection*{A relative metric: distance to the scaling laws}
In addition to overcome the nonlinearities by employing the logarithmic of the urban indicators, we may also account for the scaling behavior
between the urban indicators, homicides and the population size (Fig~\ref{fig:1}) aiming a fairer comparison between cities with different 
population sizes. We thus have proposed to evaluate the differences between the actual value of the 
urban indicators and the expected by the adjusted power law, that is,
\begin{eqnarray}
D_{Y}&=&\log_{10} Y - \langle \log_{10} Y\rangle_w\nonumber\\
 &=&\log_{10} Y - (A+\beta \log_{10} N)\,.
\label{eq:distance}
\end{eqnarray}
Note that $D_{Y}$ identifies whether a urban indicator for the given city is above ($D_{Y}>0$) or below ($D_{Y}<0$) the average scaling law 
as well as how far it is. We have also evaluated this distance for the number of homicides, that is, $D_{H}=\log_{10} H - \langle \log_{10} H\rangle_w$
(note that we are committing an abuse of terminology when denoting $D$ as a distance). This is the same idea recently proposed by Podobnik~\textit{et al.}~\cite{podobnik} for quantifying the competitiveness among countries. 

\begin{figure}[!ht]
\includegraphics[scale=0.35]{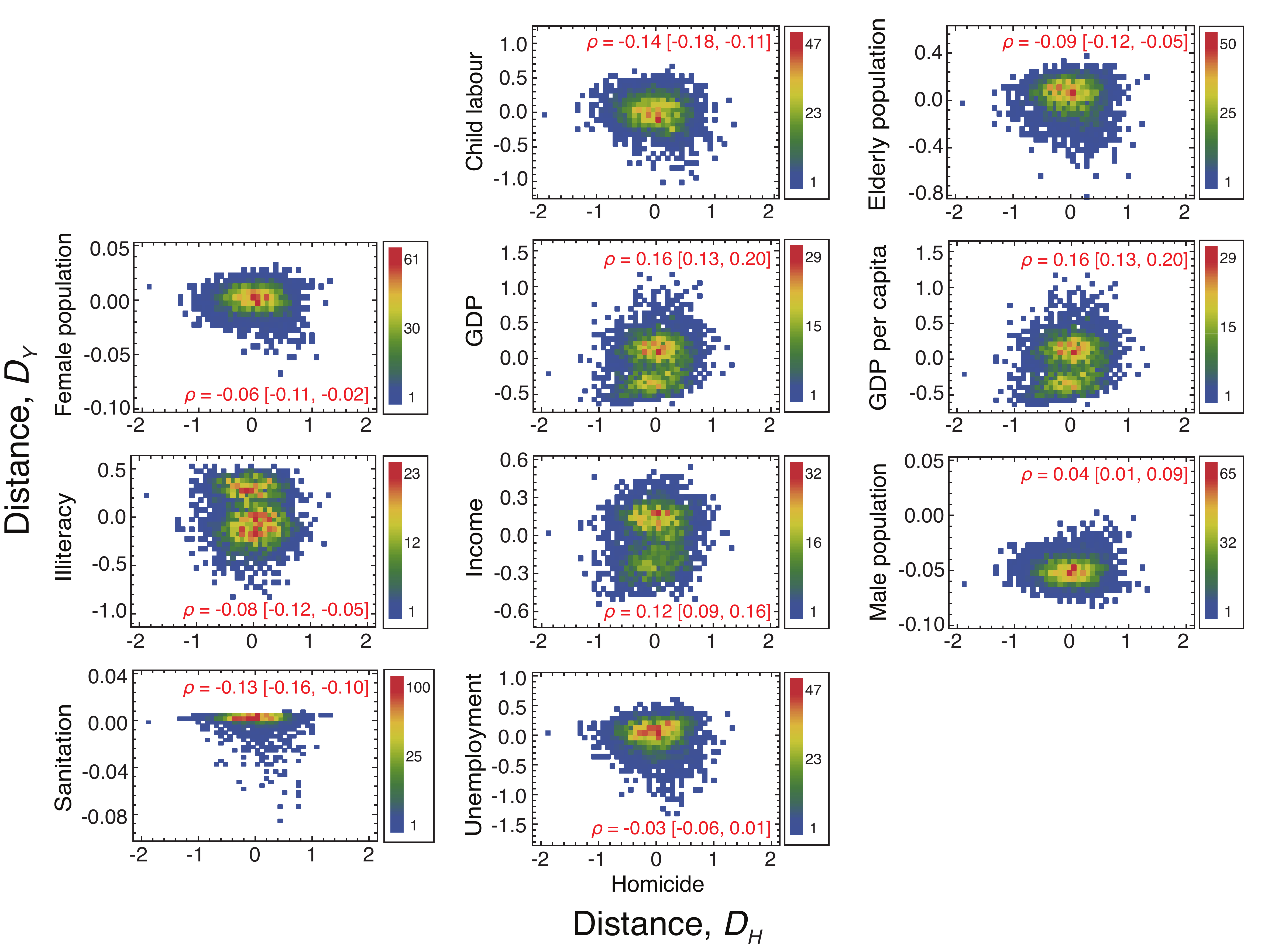}
\caption{
{\bf Distance to the scaling laws evaluated for the urban indicators versus the distance evaluated for the number of homicides.}
Scatter plot of the distances to the scaling laws evaluated for the urban indicators ($D_Y$) versus the distance evaluated for
the number of homicides ($D_H$). The color code represents the density of points, going from blue (low density) to red (hight density). We show in each plot the
value and the 95\% confidence intervals for the Pearson correlation coefficient $\rho$. We note that $D_Y$ evaluated for GDP, GDP per capita, income, and male population
are positively correlated with $D_H$, while $D_Y$ related to child labour, elderly population, female population, illiteracy, sanitation, and unemployment are
negatively correlated with $D_H$. We further observe the bimodal distributions of the relationships for GDP, GDP per capita, illiteracy, and income.
}
\label{fig:3}
\end{figure}

We have thus studied the relations between the distance evaluated from the homicide indicator ($D_{H}$) and the other urban metrics ($D_{Y}$).
Figure~\ref{fig:3} shows a scatter plot of $D_{Y}$ versus $D_{H}$, where we note that all of the urban metrics distances (except unemployment) 
have statistically significant correlations with the homicide distance (see the values of Pearson correlation $\rho$ in these plots). 
We have also observed that the sign of the correlation coefficient $\rho$ agrees with value of the linear coefficient $C_k$ for the indicators child labour, 
elderly population, female population, GDP, income, sanitation, and unemployment. However, for the indicators GDP per capita, illiteracy and male 
population, the sign of $\rho$ is opposite to the signal of $C_k$. This result means, for instance, that while the regression analysis suggests that
the increase in the male population is followed by a decrease in the number of homicides, the results when considering the relative distances point out
that the more the male population is above the power law tendency, the more the number of homicides is above the power law tendency. Similar
controversial conclusions are obtained for the indicators GDP per capita and illiteracy. 

In addition to the value of the Pearson correlation $\rho$, the scatter plots in Fig.~\ref{fig:3} reveals other intriguing patterns. We note that the
relation between the homicide distance and the indicators GDP, GDP per capita, illiteracy, and income are characterized by two peaks in 
the density of points, while for all the other indicators the density of points displays only one peak. We also note that both peaks of these 
bimodal distributions are located around $D_H\approx 0$. This result indicates that, despite the positive values of $\rho$, there is a considerable
number of cities that displays distance values for $D_Y$ above and below the power law tendency with approximately the same value for the distance $D_H$,
suggesting that such indicators may not be as good as the other ones for describing the number of homicides. 

\begin{figure}[!ht]
\begin{center}
\includegraphics[scale=0.39]{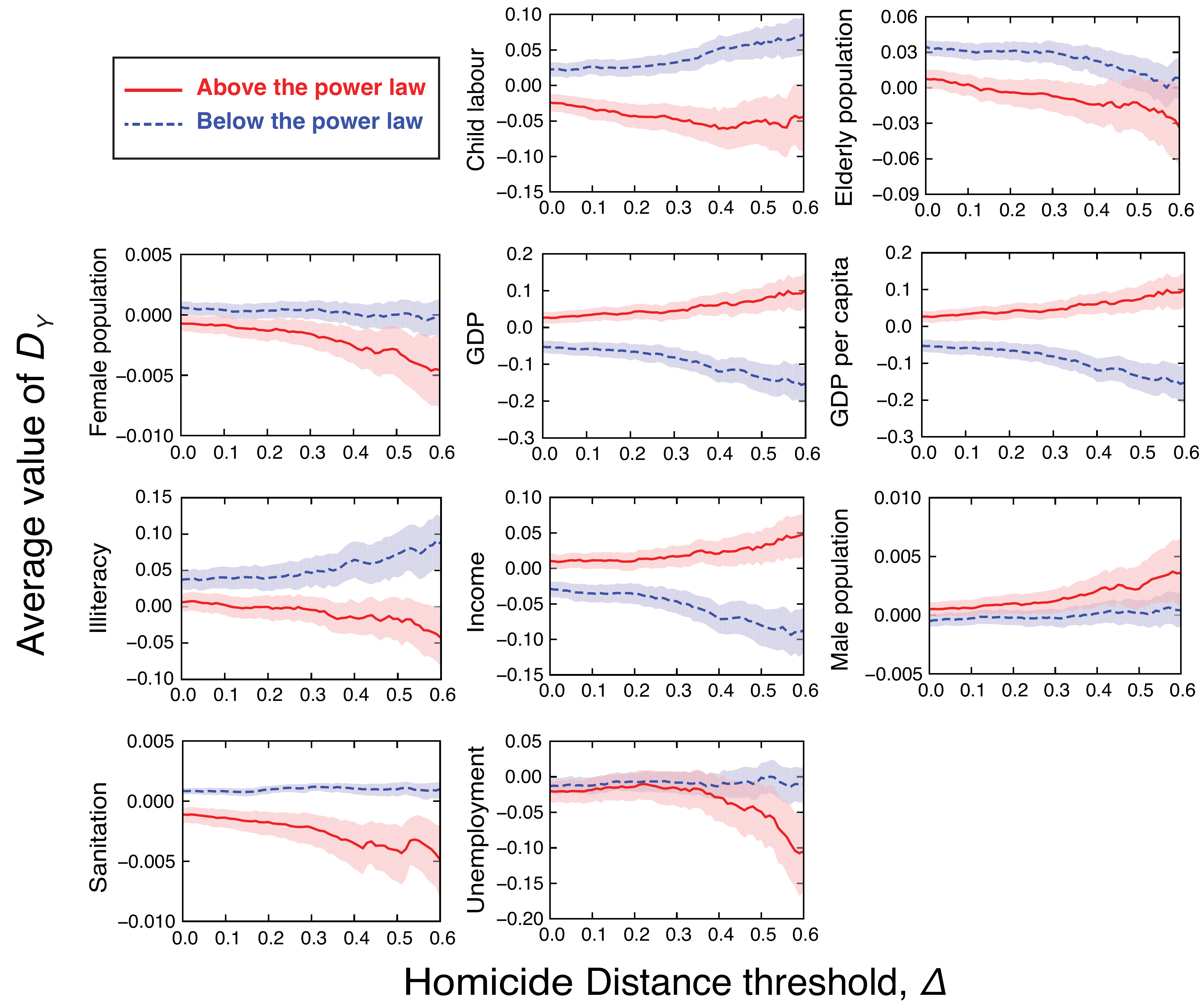}
\end{center}
\caption{
{\bf Average values of the distances to the scaling laws versus the homicide distance threshold.} The average values of distances evaluated for each urban indicator
in function of the homicide distance threshold $\Delta$, after grouping the cities that are above (red continuous lines) and below (blue dashed lines) the scaling laws with the population size. The shaded areas are 95\% confidence intervals for these average values obtained via bootstrapping~\cite{efron}.
}
\label{fig:4}
\end{figure}

Another manner of extracting meaningful information from Fig.~\ref{fig:3} is by evaluating average values. In order to do so, we have
grouped the cities in two sets: those having $D_H>0$ (homicides above the power law) and those with $D_H<0$ (homicides below the power law). We next evaluate
the average value of $D_Y$ for each group and considering the cities with absolute value of $D_H$ larger than a threshold $\Delta$. Figure~\ref{fig:4} shows these
average values as a function of the \mbox{threshold $\Delta$}. We have observed that for the indicators child labour, illiteracy and sanitation, the average
values of $D_Y$ are significantly different between the two groups of cities and also that the average of $D_Y$ increases as $\Delta$ increases 
for the cities with $D_H>0$ and decreases for those ones with $D_H<0$. The opposite occurs for the indicators GDP, GDP per capita and income, that
is, the average of $D_Y$ decreases as $\Delta$ increases for the cities with $D_H>0$ and increases for those ones with $D_H<0$. 
Intriguingly, for the indicator elderly population we observe that cities with $D_H$ below the power law present an average value of $D_Y$ larger than those with 
$D_H$ above the power law; however, this difference is only statistically significant for $\Delta\lesssim 0.45$. This result suggests that, for cities
having a much larger or much smaller number of homicides than the expected by the power law tendency, the elderly population may have no 
explanatory potential. Similarly, for the unemployment indicator, no difference is observed between the average values of $D_Y$ above and below the power law 
until $\Delta\gtrsim 0.56$. For slightly smaller value of $\Delta$, the average value of $D_Y$ (for unemployment) for cities above the power law starts to systematically
decrease and for $\Delta \approx 0.56$ a statistically significant difference is observed. This result thus provides us a clue for a better understanding
of the explicative potential of the unemployment indicator, by pointing out that (in our data) its effect is only manifested when $D_H$ is much above of the
value expected by the scaling law. 

Figure~\ref{fig:4} also provides clues of a gender effect in the number of homicides. For female population,
we note that cities with number of homicides above the power law ($D_H>0$) are characterized by an average value of $D_Y<0$ that decreases as the value of 
$\Delta$ increases. We also observe that the confidence intervals for the average values of $D_Y$ above and below the power law barely overlap each other. 
These results thus point out that in cities where the number of homicides is above the expected value, the female population is systematically smaller than
the value expected by the scaling law. For male population, despite the overlapping in the confidence intervals for the average of $D_Y$, we observe an
opposite behavior, that is, cities with number of homicides above the power law are also characterized by a male population above the power law.

\subsection*{Summary and Conclusions}
We have extensively characterized some relationships between crime and urban metrics. We have initially shown that urban indicators obey well defined
average scaling laws with the population size and also that the fluctuations around these tendencies are log-normally distributed. Using these results,
we have shown that the scaling laws can be represented by a multiplicative stochastic-like equation (Eq.~\ref{eq:powerlaw}) driven by a log-normal noise.
Next, we have addressed the problem of applying regression analysis for explaining the number of homicides $H$ in terms of urban indicators $Y$.
Because of the intrinsic nonlinearities, we have argued that it is better to employ the logarithms of these variables when performing linear regression
analysis (Eq.~\ref{eq:powerlaw} and Table~1). Furthermore, we have also discussed that accounting for the scaling phenomenon is also important for a fairer comparison among cities with different population sizes. We have thus proposed to evaluate the distances between the
actual number of homicides $H$ ($D_H$) as well as the value of the urban indicator $Y$ ($D_Y$) and the one expected by the average scaling laws. By 
investigating the Pearson correlations ($\rho$) of the relationships between $D_H$ and $D_Y$, we have found that the value of $\rho$ have the same signal of the linear coefficient
$C_k$ for the indicators child labour, elderly population, female population, GDP, income, sanitation, and unemployment. On the other hand, for 
GDP per capita, illiteracy and male population the signal of $\rho$ and $C_k$ are opposite. In addition to the values of $\rho$, we have analyzed the average
values of $D_Y$ after grouping the cities in two sets: those with number of homicides above the power law ($D_H>0$) and those below the power law ($D_H<0$).
This analysis has unveiled intriguing patterns that were not carried out by the linear regression.
In particular, our results for Brazilian cities pointed out that $i)$ the elderly population may have no explanatory potential when the number of homicides is much above or much below of the expected values by the scaling law, $ii)$ that the effect of unemployment in the number of homicides is only observed for cities with $D_H$ considerably larger than the expected by the power law, and $iii)$ that there are gender differences in the number of homicides,  
where cities with female population below the expected value  are characterized by a number of homicides above the power law and that cities with number of homicides above the power law are also characterized by a male population above the power law. We further believe that the present approach can be applied
to other datasets in order to produce more robust relationships between crime indicators and urban metrics.


\section*{Funding}
This work has been supported by the agencies Conselho Nacional de Desenvolvimento Cient\'ifico e Tecnol\'ogico (CNPq), Coordena\c{c}\~ao de 
Aperfei\c{c}oamento de Pessoal de N\'ivel Superior (CAPES) and Funda\c{c}\~ao Arauc\'aria (FAPPR). HVR is especially grateful to 
Funda\c{c}\~ao Arauc\'aria for financial support under grant number 113/2013. The funders had no role in study design, 
data collection and analysis, decision to publish, or preparation of the manuscript.

\section*{Author Contributions}
Conceived and designed the experiments: LGAA, HVR, RSM. \\
Performed the experiments: LGAA, HVR. \\
Analyzed the data: LGAA, HVR. \\
Contributed reagents/materials/analysis tools: LGAA, HVR, EKL, RSM. \\
Wrote the paper: LGAA, HVR, EKL, RSM. \\
Prepared the figures: LGAA, HVR. \\


\section*{Supplementary Information}
\subsection*{Definitions, data sources and additional comments}

We have obtained data of all Brazilian cities in the year of 2000 made free available by the Brazil's public healthcare system --- DATASUS~\cite{datasus}. Below, we describe the indicators and give some details about the data. 

\textbf{Homicide:} injuries inflicted by another person with intent to injure or kill, by any means \cite{icd}. This indicator gives the number of deaths caused by assaults. We selected the death in the DATASUS website whose cause is included in the codes X85-Y09 from the International Classification of Diseases (ICD-10) \cite{icd}.

\textbf{Population:} this indicator is derived from the population census of 2000 conducted by IBGE (Instituto Brasileiro de Geografia e Estat\'istica)~\cite{ibge} and it reports the total number of inhabitants of each city. This database also contains information about age group and gender.

\textbf{Illiteracy:} it gives the number of inhabitants of the total population in a given geographic area, in
the current year, aged 15 or older, who can not read and write at least a single ticket in the language
they know. 

\textbf{Income:} this indicator gives the average household incomes per capita of residents in a given geographic area,
in the current year. It was considered as per capita household income the sum of the monthly income of
the household, in reals divided by the number of its residents.

\textbf{Unemployment:} it gives the number of economically active population aged 16 or older who is
without work during the reference week, in a given geographic area, in the current year. It is defined as
the Economically Active Population (EAP) the number of persons aged 10 or older who are working or
looking for work. For this indicator, it was considered only the population aged 16 or older.

\textbf{Child labour:} the proportion of the population 10 to 15 years old who is working or looking for work
during the reference week, in a given geographic area, in the current year.

\textbf{GDP per capita:} Gross Domestic Product (GDP) per capita indicator is the value of the municipal GDP per capita,
being calculated as the municipal GDP of the year divided by the municipal population in the same year.
The values are presented in the currency real, not being applied deflator or no correction factor .

\textbf{GDP:} it gives the value of the municipal GDP. Values are given in thousands of the currency real,
not being applied deflator or no correction factor .

\textbf{Elderly population:} the number of inhabitant of a given city aged 60 or older.

\textbf{Sanitation:} it gives the number of inhabitants that has access to toilets, garbage collection and
water supply .

\clearpage
\setcounter{figure}{0}
\makeatletter 
\renewcommand{\thefigure}{S\@arabic\c@figure}
\renewcommand{\thetable}{S\@arabic\c@table}

\begin{figure}[!ht]
\begin{center}
\includegraphics[scale=0.43]{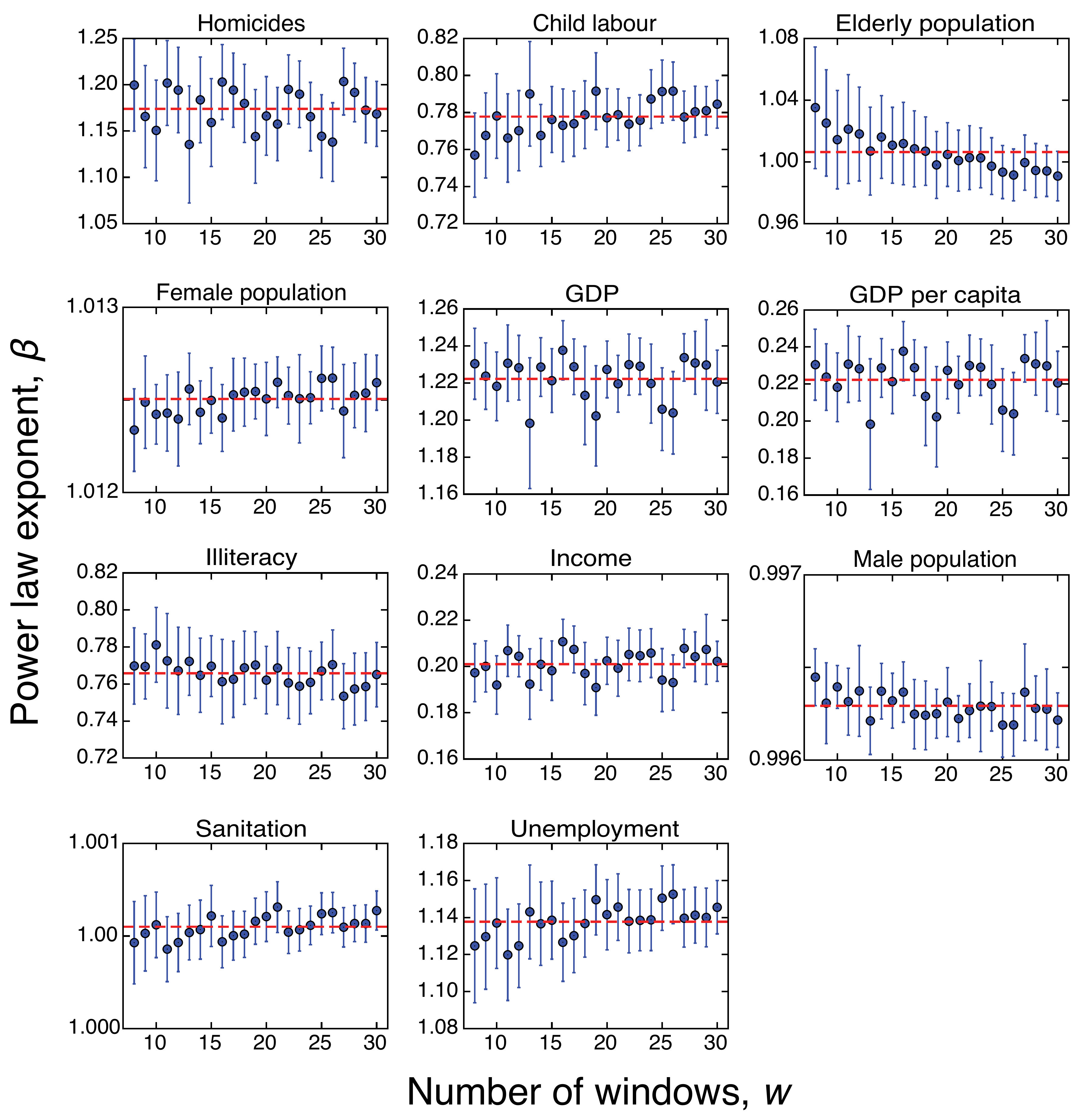}
\end{center}
\caption{
{\bf Robustness of the power law exponent versus the number of windows employed in the average relationships.} The value of power law exponent $\beta$ versus
the number of windows $w$ employed to evaluate the average relationships between $\log_{10} Y$ and $\log_{10} N$. The error bars are 95\% confidence intervals
for the value of $\beta$ and the horizontal red lines are the average values of $\beta$ over $w$. We note the almost constant behavior of $\beta$ in function of
$w$.
}
\label{sfig:1}
\end{figure}

\end{document}